\documentclass[12pt,twoside]{article}

\evensidemargin=\oddsidemargin
\def\Title#1#2#3{%
    \baselineskip=18pt
    \begin{center}
          {\Large\bf\uppercase{#1} \\ }
          \bigskip\bigskip
          {#2} \\
          {#3} \\
    \end{center}}

\long\def\Abstract#1{%
         \bigskip
         \parbox{0.93\textwidth}{%
                 \begin{center}
                       {\bf Abstract} \\
                 \end{center}
                 \medskip{\baselineskip=14pt #1}
                 \vss}
         \bigskip}

\begin{document}

\Title{GROUNDS FOR QUANTUM GEOMETRODYNAMICS \\
IN AN EXTENDED PHASE SPACE\\
AND ITS COSMOLOGICAL CONSEQUENCES}%
{T. P. Shestakova}%
{Department of Theoretical and Computing Physics, Rostov State University, \\
Sorge Str. 5, Rostov-on-Don 344090, Russia \\
E-mail: stp@phys.rnd.runnet.ru}

\Abstract{Quantum geometrodynamics (QGD) in extended phase space (EPS)
essentially distinguished from the Wheeler -- DeWitt QGD is proposed. The
grounds for constructing a new version of quantum geometrodynamics are
briefly discussed. The main part in the proposed version of QGD is given
to the Schr\"odinger equation for a wave function of the Universe.
The Schr\"odinger equation carries information about a chosen gauge
condition which fixes a reference system. The reference system is
represented by a continual medium that can be called ``the gravitational
vacuum condencate". A solution to the Schr\"odinger equation contains
information about the integrated system ``a physical object + observation
means (the gravitational vacuum condensate)". It may be demonstrated that
the gravitational vacuum condensate appears to be a cosmological evolution
factor.}

\vspace{5mm}
In this talk I briefly describe the results of the work done by
G. M. Vereshkov, V.~A. Savchenko and me\cite{Ours}. Our aim was to explore
the possibility of constructing quantum geometrodynamics of a closed
universe by a strict mathematical method without using any assumption not
permitting detailed mathematical proofs. The proposed version of quantum
geometrodynamics turns out to be a gauge-noninvariant theory {\it radically}
distinguished from the Wheeler -- DeWitt QGD by its content. The foundations
for this investigation lay in the peculiarities of the Wheeler -- DeWitt QGD
among which worth mentioning are mathematical problems of gauge invariance
and problems of interpretation.

One possible starting point for constructing QGD is to introduce the
Batalin -- Fradkin -- Vilkovisky (BFV) transition amplitude
\begin{equation}
\label{amp}
\langle\,f|i\,\rangle=\!\int\!D\mu\;\exp\,\left(iS_{eff}\right),
\end{equation}
where $S_{eff}$ is a BRST-invariant effective action, $D\mu$ is a measure
in extended phase space\cite{Hennaux}. Independence of the amplitide
(\ref{amp}) on a gauge condition is ensured by asymptotic boundary
conditions. However, there is no asymptotic states in a closed universe
and an observer cannot be removed to infinity from the investigated object
which is the Universe as a whole. So appealling to asymptotic boundary
conditions does not seem to be justified when constructing QGD of a closed
universe. As for the transition amplitude (\ref{amp}) considered
without the asymptotic boundary conditions, there is no strict mathematical
way to prove its gauge invariance. On the contrary, it may be demonstrated
by explicit calculations that the ampitude (\ref{amp}) inevitably contains
gauge-noninvariant effects (see \cite{Ours}).

Another way is based on the Wheeler -- DeWitt equation\cite{DeWitt}.
Let us emphasize that the Wheeler -- DeWitt equation is not deducible by
correct mathematical methods from a path integral or somehow
else: it can be just postulated. The principle of gauge invariance is
commonly thought to be a motivation for postulating the Wheeler -- DeWitt
equation. On the other hand, the Wheeler -- DeWitt equation is known to be
noninvariant under choice of a gauge variable, the lapse function $N$
being usually considered as such a variable\cite{HP,Halliwell}.
However, the transition to another gauge variable is formally equivalent
to imposing a new gauge condition, and vice versa. The latter reflects an
obvious fact that the choice of gauge variables and the choice of gauge
conditions have an unified interpretation: they both determine equations
for the metric components $g_{0\mu}$, fixing a reference system. So, as a
matter of fact, the parametrisation noninvariance of the Wheeler -- DeWitt
equation is ill-hidden gauge noninvariance.

It is well-known that in the Wheeler -- DeWitt theory there is no
quantum evolution of state vector in time. A wave function satisfying the
Wheeler -- DeWitt equation describes the past of the Universe
as well as its future with all observers being inside the Universe in
different stages of its evolution, and all observations to be made by these
observers. Thus the Wheeler -- DeWitt theory does not use the postulate
about the reduction of a wave packet. One cannot appeal to the Copenhagen
interpretation of quantum theory within the limits of the Wheeler -- DeWitt
QGD. One has to turn to the many-worlds interpretation of the wave function
proposed by Everett\cite{Everett} and applied to QGD by
Wheeler\cite{Wheeler}. The wave function of the Universe satisfying the
Wheeler -- DeWitt equation and certain boundary conditions is thought
to be a branch of a many-worlds wave function that corresponds to a certain
universe; other branches being selected by other boundary conditions. So
the information about the continuous reduction of the wave function in the
process of evolution of the Universe including certain observers inside
is contained in the boundary conditions for the wave function only.

Bearing in mind all the mentioned above, we propose to investigate a
more general theory. The features of the theory are:
\begin{list}{$\bullet$}{\topsep=3pt\itemsep=3pt\parsep=0pt}
\item A basic equation for a wave function of the Universe is derived by
a well-defined mathematical procedure.
\item The assumption about asymptotic states is not taken into account.
\item The theory admits of Copenhagen interpretation.
\end{list}

To illustrate our results it is convenient to consider the Bianchi IX model
for its mathematical simplicity and physical meaningfulness. I shall
remind that the interval in the Bianchi IX model looks like
\begin{equation}
\label{ds}
ds^2=N^2(t)\,dt^2-\eta_{ab}(t)e^a_ie^b_kdx^idx^k;
\end{equation}
\begin{equation}
\label{eta_ab}
\eta_{ab}(t)=\mathop{\rm diag}\nolimits\left(a^2(t),b^2(t),c^2(t)\right).
\end{equation}
We use the parametrization
\begin{equation}
\label{abc}
a=\displaystyle\frac12 r\exp\left[
 \frac12\left(\sqrt{3}\,\varphi+\chi\right)\right];\quad
b=\displaystyle\frac12 r\exp\left[
 \frac12\left(-\sqrt{3}\,\varphi+\chi\right)\right];\quad
c=\displaystyle\frac12 r\exp\left(-\chi\right);
\end{equation}
\begin{equation}
\label{zeta}
Q^a=(q,\,\varphi,\,\chi,\,\phi,\,\ldots);\quad
q = 2\ln r;\quad
\zeta(\mu,Q)=\ln\frac{r^3}N,
\end{equation}
where $\phi$ stands for scalar fields, $\zeta(\mu,Q)$ is an arbitrary
function defining a gauge variable $\mu$ through the lapse function $N$. The
Bianchi IX model can be considered as a model of a Friedman -- Robertson --
Walker closed universe with $r(t)$ being a scale factor, on which a
transversal nonlinear gravitational wave $\varphi(t),\chi(t)$ is superposed.

The derivation of an equation for a wave function of the Universe implies
going over to a path integral with the effective action in a Lagrange form.
Since the algebra of transformations generated by constraints is closed
for the model, the transition amplitude (\ref{amp}) can be reduced to the
path integral over extended configurational space involving ghost and
gauge variables with the Faddeev -- Popov effective action
$$
S_{ef\!f}=
 \!\int\!dt\,\biggl\{\displaystyle\frac12\exp\left[
   \zeta\left(\mu,Q^a\right)\right]\gamma_{ab}\dot{Q}^a\dot{Q}^b
  -\exp\left[-\zeta\left(\mu,Q^a\right)\right]U\left(Q^a\right)+
$$
\begin{equation}
\label{Seff}
  +\lambda\left(\dot{\mu}-f_{,a}\dot{Q}^a\right)
  +\frac i{\zeta_{,\mu}}\dot{\bar{\theta}}\dot{\theta}\biggr\}.
\end{equation}
We confine attention to the special class of gauges not depending on
time
\begin{equation}
\label{mu,f,k}
\mu=f(Q)+k;\quad
k={\rm const},
\end{equation}
or, in a differential form,
\begin{equation}
\label{diff_form}
\dot{\mu}=f_{,a}\dot{Q}^a,\quad
f_{,a}\stackrel{def}{=}\frac{\partial f}{\partial Q^a};
\end{equation}
$\zeta_{,\mu}=\!\partial \zeta(\mu,Q)/\partial\mu;\;\theta,\bar{\theta}$ are
the Faddeev -- Popov ghosts after replacement
$\bar{\theta}\to-i\bar{\theta}$; indices $a,b,\ldots$ are raised and lowered
with the ``metric"
\begin{equation}
\label{3-metric}
\gamma_{ab}=\mathop{\rm diag}\nolimits(-1,\,1,\,1,\,1,\,\ldots);
\end{equation}
\begin{equation}
\label{U,Ug,Us}
U(Q)=e^{2q}\,U_g(\varphi,\chi)+e^{3q}\,U_s(\phi),
\end{equation}
\vspace{-7mm}
\begin{eqnarray}
U_g(\varphi,\chi)&=&
 \frac23\left\{\exp\left[2\left(\sqrt{3}\,\varphi+\chi\right)\right]
 +\exp\left[2\left(-\sqrt{3}\,\varphi+\chi\right)\right]
 +\exp(-4\chi)-\right.\nonumber\\
&-&\left.2\exp\left[-\left(\sqrt{3}\,\varphi+\chi\right)\right]
 -2\exp\left(\sqrt{3}\,\varphi-\chi\right)
 -2\exp(2\chi)\right\}.
\end{eqnarray}

Variation of the action (\ref{Seff}) yields a Lagrangian set of equations
mathematically equivalent to canonical equations in EPS. This set of
equations can be called {\it conditionally-classical} for the presence of
Grassmannian variables. It is gauge-noninvariant.

Any gauge condition fixes a reference system, the latter representing the
observer in the theory of gravity. So the action (\ref{Seff}) describes
the integrated system ``the physical object + observation means". According
to Landau and Lifshitz\cite{Lan}, a continual medium with broken symmetry
under diffeomorphism group transformation must be considered as a reference
system in the theory of gravity. Inside the medium a periodic process is
going, its characteristic being used for choosing metric measurements
standards. Let us note that within the limits of the classical theory one
is not able to point to an object with the above properties. However,
quantum theory gives us the notion about such an object -- it is the vacuum
condecate. Thus the gauge-fixing term in (\ref{Seff}) corresponds a specific
subsystem being referred to as ``the gravitational vacuum condensate". The
investigation of the conditionally-classical set of equations reveals
the existence of a conserved quantity $E$ describing the subsystem. As
a result, the Hamiltonian constraint $H_{ph}=0$ of general relativity is
replaced by the constraint $H=E$, where $H_{ph}$ is a Hamiltonian of
gravitational and matter fields, $H$ is a Hamiltonian in EPS.

The latter means that a Hamiltonian spectrum in the appropriate quantum
theory is not limited by the unique zero eigenvalue. So the main part in
this version of QGD is given to {\it the Schr\"odinger equation in extended
configurational space} for a wave function of the Universe, and finding a
spectrum of $E$ becomes one of the main tasks of quantum geometrodynamics in
EPS. We would like to emphasize that the problem of time, so typical of the
Wheeler -- DeWitt QGD, does not arise here. Moreover, one gains the
opportunity to appeal to the Copenhagen interpretation.

The Schr\"odinger equation derived from the path integral with the effective
action (\ref{Seff}) by the standard method\cite{Cheng} originated from
Feinman reads
\begin{equation}
\label{SE1}
i\,\frac{\partial\Psi(Q^a,\mu,\theta,\bar{\theta};\,t)}{\partial t}
 =H\Psi(Q^a,\,\mu,\,\theta,\,\bar{\theta};\,t),
\end{equation}
where
\begin{equation}
\label{H}
H=-i\,\zeta_{,\mu}\frac{\partial}{\partial\theta}
  \frac{\partial}{\partial\bar{\theta}}
 -\frac1{2M}\frac{\partial}{\partial Q^{\alpha}}MG^{\alpha\beta}
  \frac{\partial}{\partial Q^{\beta}}+{\rm e}^{-\zeta}(U-V);
\end{equation}
\begin{equation}
\label{M}
M={\rm const}\cdot \zeta_{,\mu}\exp\left(\frac{K+3}2\,\zeta\right);
\end{equation}
\vspace{-4mm}
\begin{eqnarray}
V&=&-\frac3{12}\frac{(\zeta_{,\mu})^a(\zeta_{,\mu})_a}{\zeta_{,\mu}^2}
 +\frac{(\zeta_{,\mu})^a_a}{3\zeta_{,\mu}}
 +\frac{K+1}{6\zeta_{,\mu}}\,\zeta_a(\zeta_{,\mu})^a+\nonumber\\
&+&\frac1{24}\left(K^2+3K+2\right)\,\zeta_a\zeta^a+\frac{K+2}6\,\zeta_a^a;
\end{eqnarray}
\begin{equation}
\label{Galpha_beta}
\zeta_a=\frac{\partial \zeta}{\partial Q^a}
 +f_{,a}\frac{\partial \zeta}{\partial\mu};\quad
G^{\alpha\beta}={\rm e}^{-\zeta}\left(
 \begin{array}{cc}
  f_{,a}f^{,a}&f^{,a}\\
  f^{,a}&\gamma^{ab}
 \end{array}
 \right),
\end{equation}
$\alpha=(0,a),\;Q^0=\mu,\;K$ is a number of scalar fields included in the
model; the wave function is defined on extended configurational space
with the coordinates $Q^a,\,\mu,\,\theta,\,\bar{\theta}$.

It is worth noting that no ill-definite mathematical expression arises
when deriving Eq.\,(\ref{SE1}). It is due to using the nondegenarate
conditionally-classical set of equations
to approximate the path integral instead of degenerate gauge-invariant
equations. Alternatively, Eq.\,(\ref{SE1}) can be obtained from
quantum canonical equations in EPS.

The general solution to the Schr\"odinger equation (\ref{SE1}) has the
folloowing structure:
\begin{equation}
\label{time-depend.WF}
\Psi(Q^a,\,Q^0,\,\theta,\,\bar{\theta};\,t)
 =\int\Psi(Q^a)\exp(-iEt)(\bar{\theta}+i\theta)\,
  \delta(\mu-f(Q^a)-k)\,dE\,dk.
\end{equation}
where $\Psi(Q^a)$ is a solution to the stationary equation
\begin{equation}
\label{station.phys.SE}
H^0\,\Psi(Q^a)=E\,\Psi(Q^a),
\end{equation}
\begin{equation}
\label{H0}
H^0=\left.\left[-\frac1{2M}\frac{\partial}{\partial Q^a}
  M{\rm e}^{-\zeta}\gamma^{ab}\frac{\partial}{\partial Q^b}
 +{\rm e}^{-\zeta}(U-V)\right]\right|_{\mu=f(Q^a)+k}.
\end{equation}
The wave function (\ref{time-depend.WF}) carries the information on
1) a physical object, 2) observation means (a gravitational vacuum
condensate), 3) correlations between the physical object and observation
means. Observation means are represented by the factored part of the wave
function -- by the $\delta$-function of a gauge and by the ghosts; the
physical object is described by the function $\Psi(Q^a)$; the correlations
are manifested in the effective potential $V$ and in the spectrum $E$,
the gravitational vacuum condensate thus being a cosmolgical evolution
factor. The dependence of the wave function (\ref{time-depend.WF}) on
ghosts is determined by the demand of norm positivity.

The question remains if it possible to go over to some particular solution
satisfying the Wheeler -- DeWitt equation from the general solution
(\ref{time-depend.WF}). To do it, one has to eliminate correlations
between the properties of the physical object and those of observation
means. Then, one puts $E=0$ and fixes the gauge $\mu=k$, making use of
the formal possibility to go over to any given gauge by means of
transformation of the parametrization function $\zeta$. In this case
under some limitations on the measure (\ref{M})
one gets that the physical part of the wave function $\Psi(Q^a)$
satisfies the Wheeler -- DeWitt equation. However, it is of importance to
emphasize that the correlations between the physical object and observation
means cannot be eliminated completely: the information about a chosen
reference system is contained in the parametization function and the
parametrization function essentially determines the effective potential $V$.
The same picture arises when one obtains the Wheeler -- DeWitt equation as
a corollary of the superselection rules for BRST- and anti-BRST-invariant
quantum states.

In conclusion I shall describe the role of the gravitational vacuum
condensate as a cosmological evolution factor. As I mentioned above, the
gravitational vacuum condensate is a continual medium, a state equation
of the medium depending on a chosen gauge condition. The state of the
condensate is characterized by the parameter $E$, and the relation
between $E$ and other parameters of the theory determines a cosmological
scenario. For example, in a simplified model with one
of two gravitational waves, $\varphi(t)$, being frozen out, taking the
parametrization function and the gauge condition to be
$\zeta(\mu,Q^a)=\mu=k$, one obtains an ultrastiff state equation of
the condensate:
\begin{equation}
\label{state_eq}
p=\epsilon.
\end{equation}
The scalar field $\phi$ and the gravitational vacuum condensate together
form the two-component medium with a positive or negative energy density
depending on the parameter $E$. A cosmological evolution scenario may
contain the following phenomena\cite{Ours}:
\begin{list}{$\bullet$}{\topsep=3pt\itemsep=3pt\parsep=0pt}
\item cosmological expansion and contraction of space;
\item cosmological singularity;
\item compactification of space dimensions;
\item asymptotically stationary space of less dimensions;
\item inflation of the Universe.
\end{list}

The goal of the future investigation is to work out a full cosmological
scenario, in the sense as it was understood by Grishchuk and
Zeldovich\cite{GZ}, based on the proposed version of quantum
geometrodynamics. After Grishchuk and Zeldovich we think that the Universe
was created from the state ``Nothing'' where there exist neither space
with its geometry nor time. In this state there is no gravitational wave or
matter field, and no gravitational vacuum condensate either. After creation,
however, the Universe occurs in a state with broken symmetry under
diffeomorphism group, and the presence of the gravitational vacuum
condensate is a characteristic feature of this picture. A state spectrum
can be found by solving a Schr\"odinger equation for a given model.
Then, one may suppose that in the course of cosmological evolution the
Universe appears to be in the state with $E=0$, where the correlations
of physical fields with the condensate are minimal. In this state the
Universe in a large scale is described by general relativity.


\end{document}